\documentclass[aps,twocolumn,nofootinbib,superscriptaddress]{revtex4-1}
\usepackage{amssymb}
\usepackage{amsmath}
\usepackage{amstext}
\usepackage{amsfonts}
\usepackage{bbold}
\usepackage{slashed}
\usepackage{graphicx}
\usepackage{hyperref}
\usepackage[top=.7in, bottom=.7in, left=.76in, right=.76in]{geometry}

\def\l{\left}
\def\r{\right}
\def\fr{\frac}
\def\la{\label}
\def\d{\partial}

\def\be{\begin{eqnarray}}
\def\ee{\end{eqnarray}}
\newcommand{\p}{\bar{P}}  
\newcommand{\J}{\bar{J}}

\begin{document}

\title{
UV completion without symmetry restoration} 

\author{Solomon Endlich}
\affiliation{Institut de Th\'eorie des Ph\'enom\`enes Physiques, \\EPFL,
Lausanne, Switzerland}
\author{Alberto Nicolis}
\affiliation{Physics Department and Institute for Strings, Cosmology, and Astroparticle Physics,\\
  Columbia University, New York, NY 10027, USA}
\author{Riccardo Penco}
\affiliation{Physics Department and Institute for Strings, Cosmology, and Astroparticle Physics,\\
  Columbia University, New York, NY 10027, USA}

\begin{abstract}
We show that it is not  possible to UV-complete certain low-energy effective theories with spontaneously broken space-time symmetries by embedding them into linear sigma models, that is, by adding `radial' modes and restoring the broken symmetries. When such a UV completion is not possible, one can still raise the cutoff up to arbitrarily higher energies by adding fields that transform non-linearly under the broken symmetries, that is, new Goldstone bosons. However, this (partial) UV completion does not necessarily restore any of the broken symmetries. We illustrate this point by considering a concrete example in which a combination of space-time and internal symmetries is broken down to a diagonal subgroup. Along the way, we clarify a recently proposed interpretation of inverse Higgs constraints as gauge-fixing conditions.
\end{abstract}

\maketitle

\section*{Introduction}

Goldstone's theorem is one of the most powerful non-perturbative results in quantum field theory.  For Lorentz-invariant systems that exhibit spontaneous breaking of global {internal} symmetries, it provides a wealth of information by stating that their low-energy spectrum must contain one massless, spin-zero excitation---a {Goldstone boson}---for each broken symmetry. Moreover, Lorentz invariance ensures that Goldstone bosons are exactly stable,
so that they are exact eigenstates of the interacting Hamiltonian. Remarkably, all this information can be derived without making any assumption about the dynamics of the symmetry breaking mechanism. 

On the other hand, in relativistic systems where Poincar\'{e} invariance is spontaneously broken as well, the low-energy phenomenology is in general much less constrained. Goldstone bosons can have a gap\footnote{When Poincar\'{e} invariance is non-linearly realized, there is no invariant meaning of \emph{mass}. As such, we prefer to call \emph{gap} the minimum energy necessary to create an excitation.}~\cite{Nicolis:2012vf, Nicolis:2013sga, Kapustin:2012cr}, and their stability is in general no longer guaranteed by kinematics. For example, phonons in superfluid helium can be thought of as the Goldstone bosons associated with the space-time symmetries broken by the medium~\cite{Dubovsky:2005xd, Nicolis:2013lma}, and the process in which one phonon decays into two  is kinematically allowed~\cite{Endlich:2010hf}. More importantly, it was shown in~\cite{Nicolis:2013sga} that the overall {\em number} of Goldstone bosons generically depends on the dynamical details of the symmetry breaking mechanism, rather than just on the symmetry breaking pattern.

Although the situation can be very different at {low energies}, one could still conceive that most of what we know about the \emph{high-energy} behavior of ordinary Goldstones remains true even when space-time symmetries are spontaneously broken. In particular, the following three statements are usually (implicitly) regarded as indisputable in systems with spontaneously broken internal symmetries:
\begin{enumerate}
\item[(1)] The strong coupling scale of the low-energy effective theory for the Goldstones provides an estimate for the energy scale at which the symmetries are spontaneously broken.

\item[(2)] It is always possible to raise the cutoff of the low-energy effective theory by adding some \emph{radial modes}. These modes have a mass at or below the cutoff and their transformation under broken symmetries is nothing but a Goldstone-modulated unbroken transformation. In other words, if $g$ is a broken symmetry transformation, the radial modes $\psi$ transform as $\psi \to h(\pi, g) \psi$, where $h$ is an element of the unbroken symmetry group and $\pi$ stands for the collection of Goldstone modes.
This UV completion of the Goldstones' dynamics is known as a `linear sigma model': it might not be the correct description of Nature---we know that it isn't for pions---but it is always a possibility from the mathematical viewpoint.

\item[(3)] Together with the Goldstone bosons, the radial modes form a multiplet that transforms \emph{linearly} under all internal symmetries. Such a multiplet is nothing but an order parameter, whose non-zero vacuum expectation value (vev) spontaneously breaks the symmetries under consideration.\footnote{For a more precise and unambiguous characterization of {\it the} order parameter associated with a system of Goldstone fields, see \cite{Weinberg:1996kr}.}
\end{enumerate}
The main goal of this Letter is to dispel these last hopes that our intuition developed for broken internal symmetries will carry over to broken space-time ones. We do this by showing that even these three reasonable-sounding statements are not always true in the case of spontaneously broken space-time symmetries. To this end, we will consider a particularly instructive example in which a combination of internal and space-time symmetries is spontaneously broken down to a diagonal subgroup. 

Our technical analysis will be rather tedious, so it is useful to  isolate here the main features of the mechanism at work. It is well known that in the case of broken space-time symmetries, one can have fewer Goldstones than naively expected. In many cases the mismatch can be attributed to certain gauge redundancies in the Goldstone parameterization of the order parameter's fluctuations\footnote{The simplest example of this being a time-dependent scalar field, which breaks time-translations and boosts, but can only accommodate one degree of freedom.}. In other cases however---and this will be our main point---the mismatch is due to the presence of a {\em gap} for some of the Goldstones. At very low energies these can be integrated out, and one is left with an effective theory for fewer Goldstones, but with precisely the same symmetry breaking pattern---that is,  the same combination of linearly and non-linearly realized symmetries. 

Given these two options, one could further imagine that for any given symmetry breaking pattern one could realize both of these situations, depending on the precise nature of the order parameter that is breaking the symmetries.\footnote{The interested reader can find a concrete example of a similar `ambiguity' in \cite{Nicolis:2013sga}, which features two physical systems with precisely the same symmetry breaking pattern but different numbers of Goldstone excitations.} However, as we will demonstrate by example, this is not always the case since there exist symmetry breaking patterns where the first possibility is off limits. Moreover, one can tell the difference between these two scenarios already from the very low-energy viewpoint. In the latter scenario, starting from the gapless Goldstones' effective theory and going up in energy, in order to embed the effective theory into a weakly coupled UV completion, one is {\em forced} to introduce the additional Goldstones rather than a more conventional symmetry-breaking order parameter.
We dub such an unconventional UV-completion an {\em enlarged non-linear sigma model.}


\section*{Low-energy effective action}

The example we will consider 
has the Poincar\'{e} group and an internal $SO(3) \times U(1)$ symmetry 
broken down to space-time translations and spatial rotations according to the following pattern:
\be
\label{pattern}
\begin{array}{rcl}
\mbox{unbroken} &=&  \left\{
\begin{array}{ll}
\p_t \equiv P_t + \mu Q &  \quad\,\,\, \mbox{ time translations} \\
\p_i \equiv P_i &  \quad\,\,\, \mbox{ spatial translations} \\
\J_i \equiv J_i - S_i &  \quad\,\,\, \mbox{ spatial rotations}
\end{array}
\right. 
\\ && \\
\mbox{broken} &=&  \left\{
\begin{array}{ll}
K_i &  \qquad\qquad\quad\quad\,\,\,   \mbox{boosts} \\
Q &  \qquad\qquad\quad\quad\,\,\,  \mbox{internal $U(1)$} \\
S_i &  \qquad\qquad\quad\quad\,\,\,   \mbox{internal $SO(3)$} \\
\end{array}
\right. \la{pattern}
\end{array}
\ee
At first, this might seem like a very contrived symmetry breaking pattern, but in fact it is 
very similar to that characterizing  the B-phase of superfluid helium 3 \cite{PasesHe3, us}:  a broken $U(1)$ charge ($Q$) combined with an unbroken Hamiltonian at finite chemical potential ($\bar P_t$) characterizes a superfluid. Then, in helium 3's B-phase, the ``orbital'' rotations and the ``spin'' ones are broken down to a diagonal combination. Here, the spin rotations are replaced by a purely internal $SO(3)$---and the Galilei group is replaced by the Poincar\'e one. 

As a concrete example, an order parameter achieving this pattern of symmetry breaking is a complex Lorentz-vector and internal $SO(3)$-triplet, $A^a_\mu$, acquiring an expectation value
\be
\label{example_OP}
\langle A_\mu^a \rangle = f \delta_\mu^a \, e^{i \mu t} \; . 
\ee
However, without committing to any specific order parameter, one can derive the low-energy effective action describing {\em any} system characterized by the above symmetry breaking pattern using the standard coset construction~\cite{Coleman:1969sm,Callan:1969sn} for space-time symmetries~\cite{Volkov:1973vd,ogievetsky:1974ab}. The starting point is the coset parameterization
\be
\Omega = e^{i x^\mu \bar{P}_\mu}  e^{i \pi Q}  e^{i \xi^i S_i} e^{i \eta^i K_i}.
\ee
Notice that we are adopting a relativistic notation for the space-time coordinates only for reasons of typographical simplicity. Since Lorentz invariance is spontaneously broken, the $\mu=0$ and $\mu=i$ components have to be treated  independently.  

As usual, the building blocks of the low-energy effective action can be obtained by calculating the Maurer-Cartan form $\Omega^{-1} d \Omega$ and expanding its coefficients in terms of the broken and unbroken generators:
\be
\Omega^{-1} \d_\mu \Omega &\equiv& \\
&& \!\!\!\!\!\!\!\!\!\!\! i e_\mu{}^\nu ( \p_\nu + \nabla_\nu \pi Q + \nabla_\nu \xi^i S_i + \nabla_\nu \eta^i K_i  + A_\nu{}^i \bar{J}_i ). \nonumber
\ee
The quantities $ \nabla_\nu \pi, \nabla_\nu \xi^i $ and $\nabla_\nu \eta^i $ are the covariant derivatives of the Goldstone fields, while the quantity $A_\nu{}^i$ enters the effective action only at higher orders in the derivative expansion, or when couplings to `matter' fields are taken into account. 

It was first pointed out in~\cite{Ivanov:1975zq} that  when space-time symmetries are spontaneously broken, the number of Goldstone fields {\em necessary} to non-linearly realize the symmetries can be lower than naively expected.  In fact, if the pattern of symmetry breaking is such that $[ \bar{P}_a, X_b] \supset i f_{abc}Y^c$, where the $\bar{P}_a$'s are unbroken momenta and the  $X_b$'s and $Y_c$'s are broken generators that transform as multiplets under the unbroken symmetries, one can show that the covariant derivatives $\nabla_a \pi_Y^c$ contain undifferentiated $\pi_X^b$'s:
\be
\nabla_a \pi_Y^c = \partial_a \pi_Y^c - f_{abc} \pi_X^b + \mbox{higher orders}.
\ee
One then can  impose the conditions $\nabla_a \pi_Y^c \equiv 0$ to express the fields $\pi_X^b$ in terms of derivatives of $\pi_Y^c$. These conditions are known as \emph{inverse Higgs constraints}~\cite{Ivanov:1975zq} and by construction they preserve all the symmetries, even the non-linearly realized ones. In other words, by imposing some inverse Higgs constraints one obtains a non-linear realization of the same symmetry breaking pattern with fewer Goldstone fields. As pointed in \cite{Nicolis:2013sga}, whether Nature chooses to implement such constraints depends on the physical system under consideration, but from the symmetry viewpoint alone, implementing them is always a consistent possibility.


In our case, we have that $[\p_i, K_j] = - i \delta_{ij} (\p_t - \mu Q)$, and by this logic we can impose the constraint 
\be \label{nabla pi}
\nabla_i \pi = \partial_i \pi - \mu \, \eta^i + \dots = 0
\ee 
and solve it to express the  $\eta^i$ Goldstones in terms of $\pi$:
\be \la{eta}
\eta_i = \fr{\d_i \pi}{\sqrt{\d_j \pi \d^j \pi}} \, \mbox{arctanh} \l( \fr{\sqrt{\d_j \pi  \d^j \pi}}{\d_0 \pi + \mu}\r).
\ee
Notice that, despite appearances, the RHS of this equation is analytic in $\pi$.
Since the $\eta_i$'s already have one derivative per field once expressed in terms of $\pi$, we can neglect the covariant derivatives $\nabla_\mu \eta^i$ at lowest order in the derivative expansion. Therefore, the most minimal realization of the pattern of symmetry breaking (\ref{pattern}) requires only four Goldstones, namely $\pi$ and $\xi^i$, and it is described by the low-energy effective action

%
\be \la{S}
S = f^4 \int d^4 x \, \mathcal{L} (\nabla_0 \pi/f, \nabla_\mu \xi^i/f),
\ee
where 
\begin{subequations}
\be
\nabla_0 \pi &=& \Lambda_0{}^\nu (\eta) \d_\nu \pi + \mu \l[  \Lambda_0{}^0 (\eta) - \delta_\mu^0 \r] \\
\!\!\nabla_\mu \xi^j &=&\Lambda_\mu{}^\nu (\eta)\l[ \d_\nu \xi_i R^{ij} (\xi)  +  \d_\nu \eta_i  \fr{1-\cosh \eta}{\eta^2} \epsilon^{ijk}\eta_k \r] \quad\,\, \la{nablaxi}
\ee
\end{subequations}
and $\Lambda_\mu{}^\nu$ is a Lorentz boosts with rapidity $\eta^i$ given by equation (\ref{eta}) and direction such that $\Lambda_0{}^i = - \eta^i \sinh \eta / \eta$. Finally, the matrix $R^{ij} (\xi)$ is defined as
\be
R^{ij} (\xi) = \delta^{ij} + \fr{1-\cos \xi}{\xi^2} \epsilon^{ijk}\xi_k + \fr{\xi - \sin \xi}{\xi^3} (\xi^i \xi^j - \xi^2 \delta^{ij}). \nonumber
\ee

Since the unbroken symmetries include rotations, the indices of $ \nabla_\mu \xi^i$ in (\ref{S}) must be contracted 
in a manifestly {\em rotationally} invariant fashion:
this will ensure that the action is also secretly invariant under all the non-linearly realized symmetries, including Lorentz.  The scale $f$---the analog of the pion decay constant---should be thought of as the symmetry breaking scale, and in principle it does not need to be of the same order as  the scale $\mu$ \cite{Nicolis:2013sga}. 

The main reason for going through the coset construction above was to show that there is nothing pathological about the symmetry breaking pattern (\ref{pattern}). At low-energies, it leads to a well-behaved derivatively coupled theory described by the effective action (\ref{S}). For instance, by expanding this action to quadratic order in the Goldstones, one can show that all the modes are gapless and that the arbitrary coefficients in the Lagrangian can always be chosen so as to avoid instabilities.

\section*{UV completion}

We will now show that it is impossible to UV-complete the low-energy effective theory (\ref{S}) simply by adding radial modes. We demonstrate this by contradiction: If it were possible, then there would exist an order parameter $\mathcal{O}(x)$ whose non-vanishing expectation value breaks our symmetries as in
(\ref{pattern}), and whose Goldstone fluctuations  can be parameterized in terms of only four independent  modes. However, we find that the existence of such an order parameter 
conflicts with the symmetries.

Let us therefore start by assuming that such an order parameter exists. By definition, a Goldstone mode is a fluctuation of the vev of the order parameter along a direction associated with one of the broken symmetry transformations. To first order in the Goldstone fields, the most general such fluctuation can be parametrized as
\be \la{dO}
\delta \mathcal{O} (x)= i(\pi(x) Q + \eta^i (x)K_i + \xi^i (x) S_i ) \langle \mathcal{O}(x) \rangle.
\ee
If at low energies there are only four independent modes, the seven fields $\pi$, $\xi^i$, and $\eta^i$ must provide a redundant description of the low-energy fluctuations of the order parameter.  In other words, the physical fluctuation $\delta \mathcal{O}$ must be invariant under three independent gauge transformations of the fields $\pi$, $\xi^i$, and $\eta^i$ \cite{Low:2001bw, Nicolis:2013sga}. In this case, the inverse Higgs constraints (\ref{nabla pi}) should be interpreted as gauge fixing conditions that preserve all the global symmetries~\cite{Nicolis:2013sga}. 

Finding a gauge transformation
\be
\pi \to \pi + \Delta \pi, \quad \xi^i \to \xi^i + \Delta \xi^i,  \quad \eta^i \to \eta^i + \Delta \eta^i
\ee
that leaves $\delta \mathcal{O}$ invariant is equivalent to finding a non-trivial solution to the equation
\be \la{MLeq}
(\Delta \pi(x) Q + \Delta \eta^i(x) {K}_i + \Delta \xi^i(x) {J}_i  )\langle \mathcal{O}(x) \rangle \equiv 0 \; .
\ee
Notice that we were allowed to replace $S_i$ with $J_i$, since their difference is assumed to be unbroken. A criterion of this sort was first proposed in~\cite{Low:2001bw} as a way to determine the number of independent Goldstone modes. It is important to stress that whether or not equation (\ref{MLeq}) admits non-trivial solutions depends crucially on the transformation properties of $\mathcal{O}$, and not just on the pattern of symmetry breaking (\ref{pattern}) \cite{Nicolis:2013sga}. We are now going to show that in fact the only solution to equation (\ref{MLeq}) is ${\Delta \vec \xi} = {\Delta \vec \eta}= \Delta \pi = 0$, and therefore that there is no gauge transformation of the Goldstone fields that leaves $\delta \mathcal{O}$ invariant. The proof is a bit technical and it involves four main steps.\\

\noindent\emph{1. Recast the problem as an eigenvalue equation.} We start by decomposing the generators of rotations and boosts as $J_i = \epsilon_{ijk} x^j P^k + \hat{J}_i$ and $K_i = x_i P_t + t P_i + \hat{K}_i$, where $\hat{J}_i$ and $\hat{K}_i$ are finite-dimensional representations of the Lorentz group generators. The order parameter $\mathcal{O}$ does not need to transform according to an irreducible representation of $SO(3,1) \times SO(3) \times U(1)$, but without loss of generality we can restrict ourselves to this case. In fact, if we can prove that equation (\ref{MLeq}) does not have non-trivial solutions for any irreducible representation, the same will hold \emph{a fortiori} for reducible representations. For an irreducible representation we simply have $Q \langle \mathcal{O} \rangle = q \langle \mathcal{O} \rangle$, where $q$ must be non-zero because $Q$ is spontaneously broken. Using this fact, we can rewrite equation (\ref{MLeq}) as 
\be \la{MLeq3}
 (\Delta \eta^i \hat{K}_i + \Delta \xi^i \hat{J}_i )\langle \mathcal{O} \rangle = - q (\Delta \pi - \mu x^i \Delta \eta_i) \langle \mathcal{O} \rangle.
\ee
This means that, at any $x$, $\langle \mathcal{O} \rangle$ must be an eigenvector of the operator $(\Delta \eta^i \hat{K}_i + \Delta \xi^i \hat{J}_i )$ with eigenvalue $- q (\Delta \pi - \mu x^i \Delta \eta_i)$. \\

\noindent\emph{2. Show that $\Delta \eta^i \hat{K}_i$ and $\Delta \xi^i \hat{J}_i$ commute.} Since both $\bar{J}_i$ and $P_i$ are unbroken, the combination $\hat{J}_i - S_i$ must be unbroken as well, i.e.
\be \la{Jhat-S}
(\hat{J}_i - S_i)\langle \mathcal{O} \rangle = 0.
\ee
Let us therefore act with $\Delta \eta^i (\hat{J}_i - S_i )$ on both sides of equation (\ref{MLeq3}). The fact that $\Delta \eta^i, \Delta \xi^i$ and $\Delta \pi$ depend in principle on the coordinates is immaterial, because we are only considering the finite-dimensional representation of spatial rotations. Using the Poincar\'{e} algebra and equation (\ref{Jhat-S}), we get
\be
\epsilon_{ijk} \Delta \eta^i \Delta \xi^j S^k \langle \mathcal{O} \rangle = 0.
\ee
This equation seems to imply that there is a linear combination of the generators $S^k$ that remains unbroken. 
Since however the $S^k$'s are all broken by assumption, we must have that $\epsilon_{ijk} \Delta \eta^i \Delta \xi^j =0$. In particular, this means that
\be
[ \Delta \eta^i \hat{K}_i, \, \Delta \xi^i \hat{J}_i ] = i \epsilon_{ijk} \Delta \eta^i \Delta \xi^j \hat{K}^k = 0.
\ee

\noindent\emph{3. Show that  $\langle \mathcal{O} \rangle$ is an eigenvector of  $\Delta \eta^i \hat{K}_i$ and $\Delta \xi^i \hat{J}_i $.} Since the two operators $\Delta \eta^i \hat{K}_i$ and $\Delta \xi^i \hat{J}_i $ commute with each other, they must have a common basis of eigenvectors. However, this is not enough to conclude right away that $\langle \mathcal{O} \rangle$ must be separately an eigenvector of $\Delta \eta^i \hat{K}_i$ and $\Delta \xi^i \hat{J}_i$.\footnote{For instance, a singlet state $| \uparrow \, \downarrow \rangle - |  \downarrow \, \uparrow \rangle$ for two spin-$\frac12$ particles $A$ and $B$ is an eigenstate of $S_3^A + S_3^B$ without being an eigenstate of $S_3^A$ or $S_3^B$, even though $[S_3^A, S_3^B]=0$.} Crucially however, all finite-dimensional irreducible representations of the Lorentz group are not unitary, and can always be chosen in such a way that the generators $\hat{J}_i$ are hermitian but the $\hat{K}_i$'s are {\em anti}-hermitian~\cite{cornwell1984group}. This means that the eigenvalues of $\Delta \xi^i \hat{K}_i$ are either zero or purely imaginary, whereas the eigenvalues of $\Delta \xi^i \hat{J}_i$ are real\footnote{The Goldstone fields are real by construction, since they are nothing but spacetime-modulated versions of a Lie group's parameters.}. 
Since the eigenvalue on the RHS of (\ref{MLeq3}) is real, $\langle \mathcal{O} \rangle$ can only contain eigenvectors of $\Delta \xi^i \hat{K}_i$ with zero eigenvalue, and thus we must have
\begin{subequations}
\be
\Delta \eta^i \hat{K}_i \langle \mathcal{O} \rangle &=& 0 \; , \la{eigen1}\\
\Delta \xi^i \hat{J}_i \langle \mathcal{O} \rangle &=& - q (\Delta \pi - \mu x^i \Delta \eta_i) \langle \mathcal{O} \rangle \; . \la{eigen2}
\ee
\end{subequations}
\\

\noindent\emph{4. Show that $\Delta \vec \xi = \Delta \vec \eta = \Delta \pi = 0$.} Let us now act with the operator $\epsilon^{ijk} \hat n_i \Delta \eta_j (\hat{J}_k - S_k )$ on equation (\ref{eigen1}), where $\hat n$ is an arbitrary unit vector perpendicular to $\Delta \vec \eta$. We obtain
\be \la{alphaK}
| {\Delta \vec\eta} \,|^2 \,  \hat n^i \hat{K}_i  \langle \mathcal{O} \rangle = 0.
\ee
Equations (\ref{eigen1}) and (\ref{alphaK}) would be satisfied for nonzero $\Delta \vec \eta$ only if $\hat{K}_i  \langle \mathcal{O} \rangle$ vanished, but this would mean that $\langle \mathcal{O} \rangle$ is a Lorentz scalar and therefore we should have $\hat{J}_i  \langle \mathcal{O} \rangle = 0$ as well. That however would be incompatible with equation (\ref{Jhat-S}) and the fact that the $S_i$ are broken. Therefore, we conclude that it is $\Delta \vec \eta$ that vanishes. 

Similarly, if we act with $\epsilon^{ijk} \hat n_i \Delta \xi_j (\hat{J}_k - S_k )$ on (\ref{eigen2}), where $\hat n$ is now an arbitrary unit vector perpendicular to $\Delta \vec \xi$, we obtain 
\be \la{alphaJ}
| {\Delta \vec \xi} \, |^2 \, \hat n^i \hat{J}_i  \langle \mathcal{O} \rangle = 0 .
\ee
If we introduce another arbitrary unit vector $\hat m$, this time perpendicular to both $\hat n$ and $\Delta \vec \xi$, we get
\be \la{finalJstep}
0 = | {\Delta \vec \xi} \, |^2 \big[ \hat m^i \hat{J}_i \, , \,  \hat n^j \hat{J}_j \big] \langle \mathcal{O} \rangle =  | {\Delta \vec \xi} \, | \Delta \xi^i \hat{J}_i  \, \langle \mathcal{O} \rangle  ,
\ee
which---for nonzero $\Delta \vec \xi \, $---clearly implies $ \Delta \xi^i \hat{J}_i  \langle \mathcal{O} \rangle = 0$. Then, by similar logic as above we conclude that $\Delta \vec \xi=0$, and then trivially $\Delta \pi=0$.

\vspace{.4cm}

This concludes our proof that equation (\ref{MLeq}) does not admit any non-trivial solution. Therefore, there is no gauge transformation that leaves $\delta \mathcal{O}(x)$ invariant, and no order parameter that realizes the pattern of symmetry breaking (\ref{pattern}) with only four Goldstones. 

\vspace{-.3cm}

\section*{Discussion}

\vspace{-.3cm}

From a low-energy perspective, our result 
means that it is not possible to UV complete the low-energy effective action (\ref{S}) simply by adding radial modes. This result is valid, in principle, for all values of $\mu$ and $f$, but it admits a very simple explanation when $\mu \ll f$. In this case, the canonically normalized fields are $\pi_c \sim f \pi$ and $\xi^i_c \sim f \xi^i$, and the EFT becomes strongly coupled at the scale $\sqrt{\mu f}$ because all the terms with an arbitrary number of powers of $\d \pi_c / (\mu f)$ become equally important.\footnote{If the generators of unbroken rotations were simply $J_i$, then it would be possible to raise the strong coupling scale up to $f$ by tuning the coefficients in the effective action to remove all the powers of $\mu$ at the denominator (see~\cite{Nicolis:2013sga} for more details). The second term in the brackets in the covariant derivative (\ref{nablaxi}) is what makes such tuning impossible in our case, and it originates precisely from the fact that the unbroken generators of rotations are $J_i - S_i$.} Radial modes would generically come into play at the scale $f$ to restore the broken symmetries as well as unitarity, but that would be ``too late'' since unitarity is already violated at the parametrically lower scale $\sqrt{\mu f}$.

At this point, the skeptical reader might wonder whether the action (\ref{S}) could ever describe the low-energy behavior of a physical system, or equivalently whether it admits a UV completion at all. In fact, we have already introduced an order parameter, given by (\ref{example_OP}), that realizes the symmetry breaking pattern (\ref{pattern}). One can check explicitly that this order parameter does not admit any gauge invariance, in accordance with our proof.  Thus, at energies smaller than $f$ the effective action contains seven Goldstone bosons: $\pi$, the $\xi^i$'s, and the $\eta^i$'s. To lowest order in the derivative expansion their action takes the form
\be
S = f^4 \int d^4 x \, \mathcal{L} (\nabla_\mu \pi/f, \nabla_\mu \xi^i/f, \nabla_\mu \eta^i/f) \; .
\ee

However, the crucial point is that only four of these Goldstones are gapless: the boost Goldstones have generically a gap of order $\mu$. This is because the covariant derivative $\nabla_i \pi$ contains an undifferentiated $\eta^i $ (see \eqref{nabla pi}). As a result, a $\nabla_i \pi \nabla^i \pi$  term in the action contains a `mass term' of the form $\mu^2 \eta_i \eta^i$. At energies smaller than their gap, the boost Goldstones can be integrated out. At tree level, this can be achieved by solving the equations of motion to express the $\eta^i$'s in terms of $\pi$ and $\xi^i$. To lowest order in derivatives, such equations will be covariant under all the symmetries\footnote{This is because the relevant terms at this order come from the variation of the action w.r.t.~an undifferentiated $\eta^i$:
\be
\frac{\partial {\cal L}}{\partial(\nabla_j \pi)} \cdot \frac{\partial(\nabla_j \pi)}{\partial \eta^i}  = 0 \; .
\ee
The matrix on the right is invertible, since it starts as $- \mu \delta_{ij} + \dots$,
so one can identify the $\partial {\cal L}/ \partial(\nabla_j \pi)$ with the lowest order equation of motion, and that is manifestly covariant.
}, 
and thus  must  take the form (assuming parity)
\be \label{eom}
F \, \nabla_i \pi + G \, \nabla_0 \xi_i + H \, \nabla_0 \eta_i = 0 \; ,
\ee
where $F$, $G$, and $H$ are invariant functions of the covariant derivatives.
Like an inverse-Higgs constraint, such an equation can be used to eliminate the $\eta_i$'s in favor of the other Goldstones, but it  is considerably more general than the `canonical' inverse Higgs constraint \eqref{nabla pi}.
In fact, already in \cite{Ivanov:1975zq} it was pointed out that the most general inverse-Higgs constraint one should impose is a generic linear combination of covariant derivates that have, under the unbroken symmetries, the same transformation properties 
as the Goldstones one wants to eliminate. In our case this corresponds to our eq.~\eqref{eom} above with 
$F,G,H = {\rm const}$. Our analysis indicates that this is still too restrictive. At least in our weakly coupled case ($\mu \ll f$), the coefficient functions $F,G,H$ can be fairly generic invariants built out of covariant derivatives, satisfying certain integrability conditions that express that they are related to suitable derivatives of the same Lagrangian function. In practice, instead of trying to spell out and comply with these integrability conditions, it is easier to start from the action and derive the equations of motion for $\eta^i$. Once $\eta^i$ has been integrated out, the low-energy effective action one gets describes the remaining four gapless Goldstones and is precisely of the form~(\ref{S}).

This shows that the UV completion of (\ref{S}) occurs in a very unorthodox way: unitarity is restored not by radial modes, but by additional Goldstones. By including the boost Goldstones in the effective theory, we can raise the cutoff from $\sqrt{\mu f}$ to the parametrically larger scale $f$. However, the number of non-linearly realized symmetries remains the same---the UV completion does not restore any symmetry.

The particular example considered in this Letter also sheds more light on the meaning of inverse Higgs constraints: they cannot always be interpreted as gauge fixing conditions, as explicitly conjectured in~\cite{Nicolis:2013sga} (and perhaps implicitly assumed in~\cite{Low:2001bw}). As we saw, if suitably generalized, they can also correspond to integrating out at tree level and to lowest order in the derivative expansion some gapped Goldstones. In particular, we believe this to be the sense in which the results of~\cite{McArthur:2010zm} are to be interpreted. Either way, it remains true that, from a low-energy point of view, whether or not to impose inverse-Higgs constraints is always a choice: when they correspond to gauge redundancies, the choice is between inequivalent physical systems \cite{us}; when they correspond to integrating out gapped Goldstones, the choice is between working at energies much below the gap or not.

We should mention that even though  much of our discussion focused on weakly coupled UV completions (linear sigma models or enlarged non-linear ones), our proof and our considerations on the possible gauge redundancies of an order parameter are in fact more general, being independent of the weak coupling assumption: Spontaneous breaking is, by definition, characterized by order parameters. In strongly coupled theories, these can be composite operators, as is the case for the chiral condensate in QCD for instance. Yet, the Goldstone fields can still be identified with the perturbations  that are generated by applying the broken symmetries to these order parameters. Then, in the case of spontaneously broken spacetime symmetries, one can ask whether there are gauge redundancies that can affect the counting of these Goldstones. 

Finally, when $\mu$ is of the same order as the `improved' strong coupling scale $f$, it is not obvious anymore what it means to impose the available inverse Higgs constraints: The would-be gapped Goldstones of the enlarged non-linear sigma model probably have a gap of order of the strong coupling scale, making their existence as narrow resonances quite improbable, and more importantly their integrating-out quite complicated, and not just a matter of solving some lowest-order classical equation of motion.
It would be interesting to understand the physical meaning of the inverse-Higgs constraints in this case.

\vspace{.4cm}

\noindent {\bf Acknowledgments:} We are especially thankful to Rachel Rosen for collaboration during the early stages of this project. RP would like to thank Andrew Tolley, Claudia de Rham, Matteo Fasiello and Raquel Ribeiro for interesting discussions. The work of AN and RP was supported by NASA under contract NNX10AH14G and by the DOE under contract DE-FG02-11ER41743.

\bibliographystyle{apsrev4-1}
\bibliography{biblio}

\end{document}